\documentclass[twocolumn,3p,times]{elsarticle}
  
\usepackage{mathptmx, courier, pifont}
\usepackage[scaled=0.92]{helvet}
\usepackage[T1]{fontenc}
\usepackage{textcomp}
\usepackage{amsfonts,mathastext}

\usepackage{epsfig}
\usepackage{amssymb}
\usepackage{graphicx}
\usepackage{mathtools,cuted}
\biboptions{numbers,sort&compress}
\usepackage{braket}
\usepackage{physics}
\usepackage[table]{xcolor}
\usepackage{tabularx}
\usepackage{footmisc,booktabs}

\usepackage{hyperref}
\usepackage{upgreek}

\definecolor{capri}{rgb}{0.0, 0.75, 1.0}
\definecolor{cornflowerblue}{rgb}{0.39, 0.58, 0.93}
\definecolor{spirodiscoball}{rgb}{0.06, 0.75, 0.99}
\definecolor{pear}{rgb}{0.82, 0.89, 0.19}

\begin{document}

\begin{frontmatter}

\title{Extremely large oblate deformation of the first excited state in $^{12}$C:\\ a new challenge to modern nuclear theory}

\author[a]{C. Ngwetsheni}

\author[a,b]{J. N. Orce}
\ead{jnorce@uwc.ac.za}


\author[c]{P. Navr\'atil}

\author[a,d]{P. E. Garrett}

\author[e]{T. Faestermann}

\author[f]{A. Bergmaier}

\author[g]{\\M. Frosini}

\author[d]{V. Bildstein}

\author[h]{B. A. Brown}

\author[d]{C. Burbadge}

\author[i]{T. Duguet}

\author[j]{K. Hady\'nska-Kl\c ek}

\author[k,l]{\\M. Mahgoub}

\author[a]{C. V.~Mehl}

\author[g]{A. Pastore}

\author[d]{A. Radich}

\author[a]{S. Triambak}


\address[a]{Department of Physics \& Astronomy, University of the Western Cape, P/B X17, Bellville 7535, South Africa}
\address[b]{National Institute for Theoretical and Computational Sciences (NITheCS), South Africa}
\address[c]{TRIUMF, 4004 Wesbrook Mall, Vancouver, British Columbia V6T 2A3, Canada}
\address[d]{Department of Physics, University of Guelph, Guelph, Ontario N1G 2W1, Canada}
\address[e]{Physik Department, Technische Universität M\"unchen, D-85748 Garching, Germany}
\address[f]{Institute for Applied Physics and Metrology, Universit\"at der Bundeswehr M\"unchen, D-85577 Neubiberg, Germany}
\address[g]{CEA, DES, IRESNE, DER, SPRC, F-13108 Saint Paul Lez Durance, France}
\address[h]{Department of Physics \& Astronomy and National Superconducting Cyclotron Laboratory, Michigan State University, East Lansing, MI 48824-1321, USA}
\address[i]{IRFU, CEA, Universite Paris-Saclay, 91191 Gif-sur-Yvette, France}
\address[j]{Department of Physics, University of Surrey, Guildford, GU2 7XH, UK}
\address[k]{Faculty of Science, Jazan University, Jazan, Saudi Arabia}
\address[l]{Faculty of Science, Sudan University of Science and Technology, Khartoum, Sudan}

\begin{abstract}

A Coulomb-excitation study of the high-lying first excited state at 4.439 MeV in the nucleus $^{12}$C has been carried out using the $^{208}$Pb($^{12}$C,$^{12}$C$^*$)$^{208}$Pb$^*$ reaction at 56 MeV
and the {\sc Q3D} magnetic spectrograph at the Maier-Leibnitz Laboratorium in Munich.
High-statistics achieved with an average beam intensity of approximately 10$^{11}$ ions/s together with state-of-the-art {\it ab initio} calculations of the nuclear dipole polarizability
permitted the accurate determination of the spectroscopic quadrupole moment, $Q_{_S}(2_{_1}^+) = +0.076(30)$~eb, in agreement with previous measurements.
Combined with previous work, a weighted average of $Q_{_S}(2_{_1}^+) = +0.090(14)$ eb is determined, which includes the re-analysis of a similar experiment by Vermeer and collaborators,
$Q_{_S}(2_{_1}^+) = +0.103(20)$~eb.
Such a large oblate deformation challenges modern nuclear theory and emphasizes
the  need of $\alpha$ clustering and associated triaxiality effects for full convergence of $E2$ collective properties.
\end{abstract}


\end{frontmatter}




\section{Introduction}

Advanced beam delivery, high-resolution and high-efficiency detectors in modern accelerator facilities permit the
precise determination of electromagnetic properties involving high-lying excited states in light nuclei. Ergo the feasibility
to benchmark  modern theory from {\it ab initio}~\cite{entem2015peripheral,calci2016sensitivities,epelbaum2012structure,dreyfuss2013hoyle},
beyond mean-field~\cite{bender2003self,ebran2012atomic,cui2017beyond} and cluster-model~\cite{kanada2007structure,chernykh2007structure,freer2018microscopic,marin2014evidence}
 calculations.
A suitable testing ground concerns the nucleus $^{12}$C,
with a first excitation  of angular momentum $J^{\pi}=2_{_1}^+$ lying at 4.439 MeV.

Unlike excitation energies, electromagnetic transitions provide a more stringent test of the many-body Hamiltonian because they involve  the overlap between initial and final wave functions.
Particularly, the reduced  transition probability  between the ground 0$^+_{_1}$ and first-excited 2$_{_1}^+$ states
--- the  $B(E2; 0^+_{_1} \rightarrow 2_{_1}^+)=\lvert \langle 2^+_{_1}\mid\mid \hat{E2} \mid\mid 0^+_{_1}\rangle\lvert^2$ value ---
quantifies the collective motion through the $\hat{E2}$ electric-quadrupole operator.
This is precisely measured in $^{12}$C, $B(E2; 0^+_{_1} \rightarrow 2_{_1}^+)=0.00390(8)$ e$^2$b$^2$, with a remarkable  2\%   uncertainty~\cite{pritychenko2016tables,d2020precision,helm1956inelastic,strehl1970untersuchung,crannell1967determination,crannell1964determination,john2003isoscalar,kuronen1988electronic,LUXITING1988909,be2Ref10,cockburn1970lifetime,riess1968lifetimes,catz1967study,warburton1966lifetime,devons1962electromagnetic,fagot1971nearly,nndcbe2}
in agreement with state-of-the-art {\it ab initio} calculations using large harmonic-oscillator basis sizes,
$N_{{max}}$~\cite{d2020precision,navratil2000properties,forssen2013systematics}.
Albeit general concordance, Monte Carlo lattice calculations at leading order ({\sc LO}) underestimate the experimental value~\cite{epelbaum2012structure},
emphasizing the need of higher-power counting orders in chiral effective field theory ($\chi$EFT), i.e., the inclusion of three-nucleon ({\sc 3N}) forces and other non-local contributions to the nucleon-nucleon ({\sc NN}) interaction~\cite{machleidt2011chiral}.


The spectroscopic quadrupole moment of the first 2$^+_{_1}$ state, $Q_{_S}(2_{_1}^+)$,  is related to the diagonal matrix element of the $ \hat{E2}$ irreducible rank-2 tensor as
 \begin{eqnarray}\label{eq:2}
 Q_{_S}(2_1^+)
 &=&\sqrt{\frac{16\pi}{5}}\frac{1}{\sqrt{2J+1}}\langle  JJ20
\mid JJ \rangle \langle 2^+_{_1}\mid\mid \hat{E2} \mid\mid 2^+_{_1}\rangle \nonumber \\
&=&0.75793\langle 2^+_{_1}\mid\mid \hat{E2} \mid\mid 2^+_{_1}\rangle,
\end{eqnarray}
and quantifies the ellipsoidal deformation of the charge distribution in the laboratory frame~\cite{alder1975north}. 
This value is known to a much lesser extent in $^{12}$C with only
three measurements being performed ~\cite{raju2018reorientation,vermeer1983electric,saiz2021coulomb}. The $Q_{_S}(2_{_1}^+)$ value can be determined at  energies well below the Coulomb barrier for states with $J \neq 0, 1/2$ using the reorientation effect ({\sc RE}), which is a second-order effect in Coulomb-excitation perturbation theory
that causes a distinct population of the magnetic substates depending on the deformation~\cite{hausser1974coulomb}.
For the excitation probability of a $0^+_{_1}\rightarrow 2^+_{_1}$ transition in the projectile nucleus~\cite{alder1975north},
\begin{equation}
 P_{_{02}}^{(2)} = P_{_{02}}^{(1)} + P_{_{02}}^{(1,2)},
\end{equation}
where {\sc RE} is accounted by the interference term $P_{_{02}}^{(1,2)}$, which includes both E1 and E2 phenomena, adding to the first-order term $P_{_{02}}^{(1)}  \propto B(E2; 0^+_{_1} \rightarrow 2_{_1}^+)$. The contribution of the {\sc RE} in $P_{_{02}}^{(2)}$, $r_{_P}(\xi,\vartheta)$,
is given by ~\cite{deBoer1968},
\begin{equation}
r_{_P}(\xi,\vartheta) = \frac{P_{_{02}}^{(1,2)}(E2)}{P_{_{02}}^{(1)}} \approx 1.32~\frac{A_{_P}}{Z_{_P}}\frac{\Delta E}{1+A_{_P}/A_{_T}} Q_{_S}(2_{_1}^+) ~K(\xi,\vartheta),
\label{eq:re}
\end{equation}
where $\Delta E$ is the 2$_{_1}^+$ excitation energy, $Z_{_P}$ and $A_{_P}$ the charge and mass numbers of the projectile, respectively, and $K(\xi,\vartheta)$ a function, with values
ranging between 0.2 and 0.3 for this experiment~\cite{deBoer1968}, that depends on the scattering angle in the center-of-mass frame, $\vartheta$, and the adiabaticity parameter, $\xi$,
given by the ratio between the collision time, $\tau_{_{collision}}$, and the lifetime of the level $\tau_{_{level}}=\hbar/\Delta E$~\cite{deBoer1968}.
For sudden-impact collisions, $\xi \lesssim 1$,  while for adiabatic collisions, $\xi\gg1$.
A larger sensitivity to the {\sc RE} is attained for large $\big|Q_{_S}(2_{_1}^+)\big|$ values, large scattering angles where $K(\xi,\vartheta)$ is maximized, and heavy targets, $A_{_T}$~\cite{nakai1970quadrupole}.


\begin{figure*}[!ht]
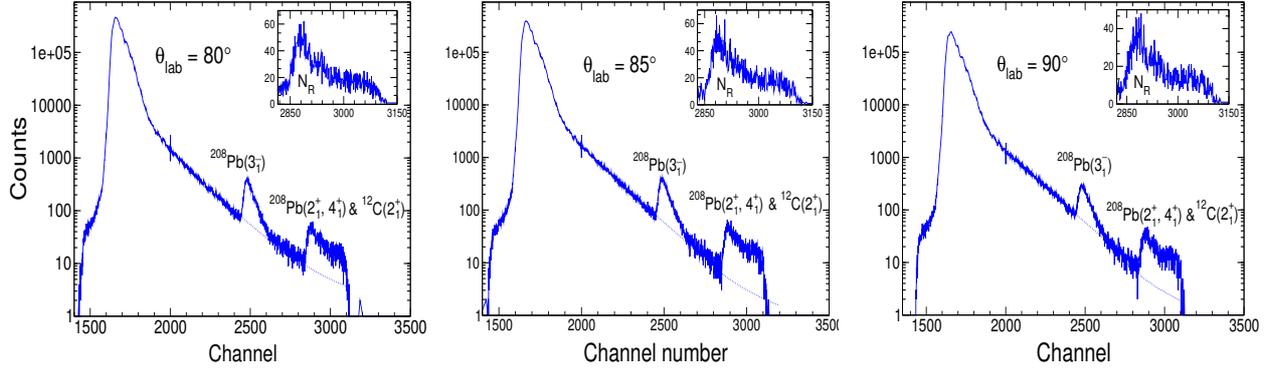

    \centering
    \includegraphics[width=5.5cm,height=4.8cm,angle=-0]{80deg_spectrum2.eps}
    \hspace{0.05cm}
    \includegraphics[width=5.25cm,height=4.8cm,angle=-0]{85deg_spectrum2.eps}
    \hspace{0.05cm}
    \includegraphics[width=5.25cm,height=4.8cm,angle=-0]{90deg_spectrum2.eps}

    \caption{Particle energy spectra collected at 80$^\circ$ (left), 85$^\circ$ (middle) and 90$^\circ$ (right) scattering angles with 56 MeV $^{12}$C ions. All identified peaks are labelled and the blue dotted line represents the elastic peak line-shape. The inset  shows an expanded spectrum with the broadened 2$^+_{_1}$ peak in  $^{12}$C overlapping with the 2$^+_{_1}$ and 4$^+_{_1}$ peaks in $^{208}$Pb.}

\label{fig1:spectrum}
\end{figure*}

An interesting correlation between $B(E2; 0^+_{_1} \rightarrow 2_{_1}^+)$ and $Q_{_S}(2_{_1}^+)$ values is inferred from {\it ab initio} calculations in $^{12}$C~\cite{calci2016sensitivities,d2020precision},
where the extrapolated $Q_{_S}(2_{_1}^+)$ value can be calculated precisely from the well-known $B(E2; 0^+_{_1} \rightarrow 2_{_1}^+)$~\cite{d2020precision}.
The largest basis size was $N_{_{max}} = 8$~\cite{calci2016sensitivities,forssen2013systematics}, where the importance-truncation No-Core Shell Model ({\sc IT- NCSM})~\cite{roth2009importance}
was used with the CD-Bonn 2000 {\sc NN} potential~\cite{machleidt2001high} and a variety of {\sc 2N+3N} chiral interactions~\cite{calci2016sensitivities}.
Values of $Q_{_S}(2_{_1}^+)$ ranging from 0.045 to 0.065 eb have been computed
from the $B(E2; 0^+_{_1} \rightarrow 2_{_1}^+)$ --- $Q_{_S}(2_{_1}^+)$ correlation~\cite{calci2016sensitivities,d2020precision}.
A similar result of  $Q_{_S}(2_{_1}^+)\approx0.06$ eb is  determined by other {\it ab initio}~\cite{epelbaum2012structure,dreyfuss2013hoyle,forssen2013systematics,d2020precision} and
theoretical approaches~\cite{abgrall1972multipole,bassel19820+}, including
shell-model ({\sc SM}) calculations using the Cohen-Kurath
interaction~\cite{cohen1965effective}
and admixtures outside the $p-$shell from the isoscalar giant quadrupole resonance~\cite{bassel19820+}.
Lastly, Monte Carlo lattice calculations~\cite{epelbaum2012structure} and cluster  models~\cite{marin2014evidence,freer2018microscopic,kanada2007structure,chernykh2007structure} predict an oblate shaped 2$_{_1}^+$ state in $^{12}$C, with a compact triangular three-$\alpha$ cluster and mean-field admixtures.
Larger $Q_{_S}(2_{_1}^+)$ values can certainly be calculated by the shell model using larger effective charges~\cite{yuan2012,kitagawa1999shell,sagawa2004deformations},
as recently reviewed by Saiz and collaborators~\cite{saiz2021coulomb}.
In this Letter we present a high-statistics {\sc RE} measurement yielding the most accurate $Q_{_S}(2_{_1}^+)$ value in $^{12}$C in agreement with previous
Coulomb-excitation measurements~\cite{vermeer1983electric,raju2018reorientation,saiz2021coulomb}. When combined with prior work,
an extremely large oblate deformation is determined in conflict with modern theoretical approaches.

\section{Experiment Details}

A safe Coulomb-excitation of the $2_{_1}^+$ state in $^{12}$C has been performed using the high resolution {\sc Q3D} magnetic spectrograph at the  Maier-Leibnitz-Laboratory (MLL) in Garching, Munich (Germany)~\cite{dollinger2018physics}.
Beams of $^{12}$C with an intensity of up to 10 pnA  with a ${5^+}$ charge state were accelerated to $T_{_{lab}}=56$ MeV by the  Tandem Van de Graaff accelerator. A 99.36\% enriched $^{208}$Pb target with 100 $\upmu g/cm^2$ thickness was located at
the object focus of the {\sc Q3D} magnetic spectrograph~\cite{dollinger2018physics}.
The scattered $^{12}$C ions were momentum analyzed by the spectrograph, with the magnetic field
such that $^{12}$C  ions corresponding to the excitation of  the high-lying states $^{12}$C($2_{_1}^+$) and
$^{208}$Pb($4_{_1}^+, 2_{_1}^+,3_{_1}^-$) states at 4.32~MeV, 4.09~MeV and 2.61~MeV, respectively,
together with those elastically scattered reached the focal plane.


Reaction products enter the {\sc Q3D} spectrograph through an aperture with a solid angle $\Omega =$ 14~msr and are focused vertically by the quadrupole magnet.
The horizontal focusing onto the focal-plane detector is achieved by the dipole magnetic field, which separates the trajectories according
to the ratio momentum/charge. 
The detector system is a combination of two aligned, 1-m long, gas filled proportional counters for position measurement using the signals from both sides of the wires
and particle identification using energy losses $\Delta E$ and a 1-m long Frisch-grid ionization chamber with an anode which has five strips,
where five  $\Delta E$ plus the total energy (E) are measured ~\cite{mayer1981}.

Data acquisition was carried out using the {\sc ROOT} based {\sc MARaBOU} software~\cite{marabou}. The resulting particle spectra obtained at
spectrograph angles of $\theta=80^{\circ}$ (left), $85^{\circ}$ (middle) and $90^{\circ}$ (right) are shown in Fig.~\ref{fig1:spectrum}.
The spectra show the dominant elastic peak, prominent 3$_{_1}^-$ peak in $^{208}$Pb and the broadened 2$_{_1}^+$ peak in $^{12}$C,
coinciding with the 2$_{_1}^+$ and 4$_{_1}^+$ peaks in $^{208}$Pb (insets in Fig.~\ref{fig1:spectrum}).
The broadening of the peak due to the $^{12}$C 2$_{_1}^+$ state  results from
the emission of the 4.439 MeV $\gamma$ ray while the $^{12}$C ions are in flight~\cite{beene1976particle}.
A similar particle spectrum was collected by Vermeer and collaborators~\cite{vermeer1983electric},
using a split-pole magnetic spectrograph, where scattered ions from $T_{_{lab}}= 53$ to $58$ MeV were detected at 90$^{\circ}$ by a position-sensitive, multi-element proportional counter. 

The semi-classical coupled-channel Coulomb-excitation code {\sc GOSIA}~\cite{czosnyka1983d} was used to simulate the corresponding population probabilities, where the semi-classical approximation
({\sc SCA}) was satisfied with a Sommerfeld parameter of $\eta = \frac{a}{\lambdabar}=36\gg 1$,
defined by the ratio between half the distance of closest approach in a head-on collision, $a$,  and the de Broglie wavelength $\lambdabar$~\cite{alder1975north}. Concurring results shown below indicate that nuclear effects are negligible at  $\theta = 80^{\circ}$, $85^{\circ}$ and $90^{\circ}$, which correspond to separations between nuclear surfaces of $S(\vartheta)=6.5$, $6.0$ and $5.6$ fm, respectively, where,
\begin{eqnarray}
S(\vartheta)&=&   \frac{1.44~ Z_{_P} Z_{_T}}{2T_{_{lab}}} (1+A_{_P}/A_{_T})\left[1+cosec(\vartheta/2)\right] \nonumber \\ &-& 1.25(A_{_P}^{1/3}+A_{_T}^{1/3})~\mbox{fm},
\end{eqnarray}
The {\sc SCA} is further supported by a similar {\sc RE} study~\cite{vermeer1983electric}, where the safe bombarding energy was measured at $90^{\circ}$, finding  consistent results at $T_{_{lab}}=53$, 54 and 56 MeV, with an onset of nuclear interference at 58 MeV~\cite{vermeer1983electric}.

\section{Data Analysis}


The extraction of the $Q_{_S}(2_{_1}^+)$ value depends on the excitation probabilities from experimental data and those calculated by {\sc GOSIA}~\cite{czosnyka1983d}. Prior determination of the experimental excitation probabilities requires background subtraction of the elastically scattered $^{12}$C ions. Here, the background subtraction was done using the line shape of the elastic peak, denoted by the fine dotted line in Fig.~\ref{fig1:spectrum}.  The line shape can be fitted with an exponential or a Landau  function, and the best fit is selected based on the minimum mean square error. Spreading of the $^{208}\mbox{Pb}(3_{1}^-)$ peak may affect the elastic peak line shape and ultimately the background subtraction of the broadened peak area $N_R$ (insets in Fig.~\ref{fig1:spectrum}). This was investigated by fitting the $^{208}\mbox{Pb}(3_{1}^-)$ peak with a Landau function extending to the peripheral of $N_R$ and compared with the exponential function fitting the elastic peak line shape, resulting in a difference of 1.8\% at 80$^{\circ}$, 7.8\% at 85$^{\circ}$
and 5.6\% at 90$^{\circ}$ in the statistics of $N_R$ after background subtraction, which is included in the uncertainties of $N_R$. The peak counts after background subtraction are then used to calculate the experimental population probabilities, i.e.,
\begin{equation}
 P(^{12}C;2_{_1}^+) = \frac{ N(^{12}C;2_{_1}^+)}{N_{total}},
 \label{eq:P_exp}
\end{equation}
where $N_{total}$ is the total number of counts recorded, including $N(^{12}C;2_{_1}^+)$.

The region $N_R$ includes the peak due to the $^{12}$C 2$_{_1}^+$ as well as contributions from the excited states in
$^{208}$Pb, namely the 2$_{_1}^+$ and 4$_{_1}^+$ states. These latter contributions must be subtracted, and this was
accomplished using the {\sc GOSIA} calculated population probabilities, $P(^{208}\mbox{Pb};2_{_1}^+,4_{_1}^+)$, as $N(^{12}C;2_{_1}^+) = N_R - N_{total}P(^{208}\mbox{Pb};2_{_1}^+,4_{_1}^+)$. Table~\ref{tbl:statistics} shows statistics and populations of the $^{12}$C$(2_{_1}^+)$ state, where the reported uncertainties include statistical and systematic uncertainties from the background subtraction procedure. We also investigated the impact of
the $E1$ transition $\langle 3^-_1 \mid\mid \hat{E1} \mid\mid 2^+_{_1} \rangle$ on P$(^{208}\mbox{Pb};2_{_1}^+)$, which is negligible $\approx 0.3\%$.
\begin{figure}[!ht]
 	\begin{center}
 	\includegraphics[width=7.5cm,height=6.cm,angle=-0]{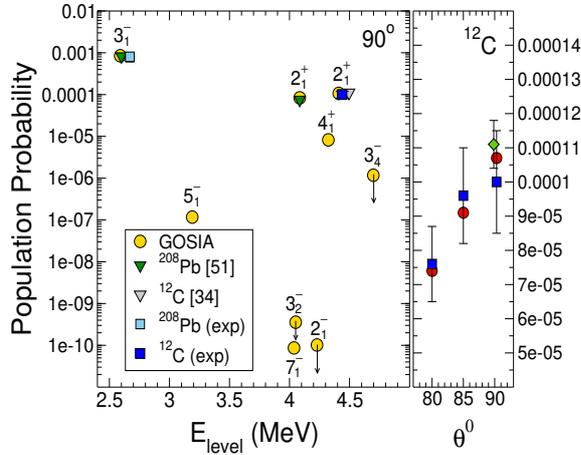}
 		\caption{(Left panel) Calculated (circles) and measured (squares) population probabilities on a logarithmic scale for $^{208}$Pb states and $^{12}$C(2$^+_{_1}$), respectively, at 90$^\circ$. The right panel also shows calculated (circles) and measured (squares) population probabilities on a linear scale for $^{12}$C versus scattering angle $\theta$. The previously measured population probabilities of $^{12}$C(2$^+_{_1}$) \cite{vermeer1983electric} and $^{208}$Pb(3$_{1}^-$, 2$_{1}^+$) states~\cite{vermeer1984Pbmeasurements} (triangles down) are shown for comparison.}
 	 		\label{fig2:probabilities}
 	\end{center}
 \end{figure}

Figure~\ref{fig2:probabilities} presents the population probabilities of all levels in the energy range of interest for both $^{12}$C and $^{208}$Pb.
Our experimental results (squares) are in agreement with preceding measurements by Vermeer and co-workers at 90$^{\circ}$ (triangles down)~\cite{vermeer1983electric,vermeer1984Pbmeasurements}
for the 2$^+_{_1}$ and 3$^-_1$ states in $^{12}$C and $^{208}$Pb, respectively. A value of $B(E3; 0_{_1}^+\rightarrow 3_{_1}^-)=0.583(18)$ e$^2$ b$^3$ has been determined for $^{208}$Pb
in agreement with previous work~\cite{goutte1980determination,spear1983improved}.
The arrows pointing down in Fig.~\ref{fig2:probabilities} are upper limits assuming a dominant $E2$ transition, where there is likely $M1/E2$ admixtures.
The population of $^{208}\mbox{Pb}(3_{_2}^-, 3_{_4}^-)$ states may contribute to the counts in the region $N_R$, and must also be accounted for.
Both states contribute insignificantly to $N_R$,
the largest being the $3_{_4}^-$ state with a maximum of  48 counts.
The resulting population of the $^{12}$C(2$_{_1}^+$) state has a sensitivity to both the $0_{_1}^+ \rightarrow 2_{_1}^+$
transition and the 2$^+_{_1}$ diagonal matrix elements, but also is sensitive to the polarizability of the $^{12}$C states.
This polarizability must be accurately accounted for in the excitation process.
\begin{table}[!ht]
\centering
\begin{tabular}{c | c c c}
    \hline
    \hline
                                             & 80${^\circ}$    & 85${^{\circ}}$       &  90${^{\circ}}$   \\
    \hline 
     N($^{12}$C;~2$_{_1}^+$)                 & 2837(412)    & 3157(447)         & 2157(320)   \\
    \hline
    10$^5$ $\times$ P($^{12}$C;~2$_{_1}^+$)  & 7.6(1.1)        & 9.6(1.4)            &   10.0(1.5)   \\
    \hline\hline
\end{tabular}
\caption{The statistics of the $^{12}$C 2$^+_{_1}$ peak and populations.  
}

\label{tbl:statistics}
\end{table}


\section{Dipole Polarizability}

A competing phenomenon in the determination of $Q_{_S}(2_{_1}^+)$ values in light nuclei is the $E1$-polarizability, which must be corrected for in Coulomb-excitation codes~\cite{czosnyka1983d}. The {\sc E1} polarization correction arises from virtual excitations of giant dipole resonance ({\sc GDR}) states,
which polarize the final excited level via the two-step process:  $\rvert i\rangle \rightarrow \rvert n\rangle \rightarrow \rvert f\rangle $ or $0_{_1}^+ \rightarrow 1_{_{GDR}}^-\rightarrow 2_{_1}^+$, for the general case of an even-even nucleus. Previous studies show that $E1$-polarizability corrections are necessary in order to reproduce the excitation probabilities measured for the $J = 1/2$ first excited levels in $^{7}$Li and $^{17}  $O~\cite{hausser1973e1,hausser1972nuclear,kuehner1982measurement}, where $Q_{_S}(J=1/2)=0$ and the $E1$ polarizability is the only second-order effect. This is included in the effective quadrupole interaction by the dimensionless polarizability parameter $\kappa$, which measures deviations between the
actual effects from the {\sc GDR} and the hydrodynamic model~\cite{orce2015new}. The value of $\kappa(0^+_{_1})$ can be determined from the $(-2)$ moment of the total photo-absorption cross section, $\sigma_{_{-2}}$~\cite{orce2015new,levinger1957migdal},
\begin{equation}
 \sigma_{_{-2}} = \int_{0}^{E_{\gamma}} \frac{\sigma_{_{total}}(E_{_{\gamma}})}{E_{_{\gamma}}^{^2}}~dE_{_{\gamma}} = 2.38 \kappa ~ A^{5/3} ~\mbox{$\upmu$b/MeV}.
\label{eq:238}
\end{equation}


As mentioned before, the interference term $P_{_{02}}^{(1,2)}$ entails an $E1$ component. Similar to Eq.~\ref{eq:re}, the GDR effect can be approximated by~\cite{deBoer1968} 
\begin{equation}
 d_p(\xi,\vartheta) = \frac{P_{_{02}}^{(1,2)}(E1)}{P_{_{02}}^{(1)}} \approx -7.37\frac{ T_{lab} \sigma_{_{-2}}  }{(1+A_p/A_t)Z_tZ_tR^2} \phi(\xi,\vartheta),
\end{equation}
where $\phi(\xi,\vartheta)$ also contains kinematical information and varies between approximately 0.03 and 0.1~\cite{deBoer1968} and $R=1.25~A^{1/3}$ fm is the spherical nuclear radius. Ultimately, we can approximate the second-order excitation probability by the sum, $P_{_{02}}^{(2)} =  P_{_{02}}^{(1)}[1+r_{p}(\xi,\vartheta) + d_p(\xi,\vartheta) ]$.
However, there is currently no experimental data of $\sigma_{_{-2}}$ or $\kappa$ values for excited states in even-even nuclei.
Recent developments using the formalism of H\"ausser and collaborators~\cite{hausser1973e1} based on $E1$ and $E2$ matrix elements,
second-order perturbation theory and the hydrodynamic model,
show that $\sigma_{_{-2}}(2^+_{_1})$ and $\kappa(2^+_{_1})$ can be calculated as~\cite{Orce2024}
\begin{eqnarray}
\sigma_{-2}(2^+_{_1}) &=& 0.6\frac{Z A^{2/3} S(E1)}{\langle 0^+_{_1}\mid\mid \hat{E2} \mid\mid 2^+_{_1}\rangle} ~~\mbox{fm$^2/$MeV}, \label{eq:extd_kappa} \\
\kappa(2^+_{_1}) &=& \frac{1}{0.00039\frac{A}{Z}}\frac{S(E1)}{\langle 0^+_{_1}\mid\mid \hat{E2} \mid\mid 2^+_{_1}\rangle}, \label{eq:extd_kappa2}
\end{eqnarray}
for the two-step process, $0_{_1}^+ \rightarrow 1_{_{GDR}}^-\rightarrow 2_{_1}^+$,
where Eqs.~\ref{eq:extd_kappa} and \ref{eq:extd_kappa2} satisfy Eq.~\ref{eq:238} and $S(E1)$ is defined as
\begin{equation}
 S(E1):= \frac{1}{2J_i+1}
 \sum_{J_n,\Delta T} W_{inf} \frac{\langle i\parallel\hat{E1}\parallel n\rangle \langle n\parallel\hat{E1}\parallel f\rangle}{E_{_{\gamma}}},
 \label{eq:SE1}
\end{equation}
in units of e$^2$fm$^2/$MeV,  where the sum considers all intermediate states $\rvert n\rangle = \rvert 1^-_{_{GDR}}\rangle$ connecting both the initial
$\rvert i \rangle = \rvert 0^+_{_1}\rangle$ and final $\rvert f \rangle =\rvert 2^+_{_1}\rangle$ states via isovector $E1$ transitions --- following the isospin selection rule $\Delta T = 1$ for $N=Z$ self-conjugate nuclei~\cite{isospinselection} ---
and $W_{inf}=W(1102,21)$ is the corresponding Racah W-coefficient~\cite{racah1942theory}.
Consequently, $\sigma_{_{-2}}(2^+_{_1})$ and $\kappa(2^+_{_1})$ values can be determined from theoretical models by computing $E1$ and $E2$ matrix elements.
The calculated $\kappa(2^+_{_1})$ value is then input in {\sc GOSIA}~\cite{czosnyka1983d} for further analysis.

\begin{figure}[!ht]
\begin{center}
\includegraphics[width=7.6cm,height=6.6cm,angle=-0]{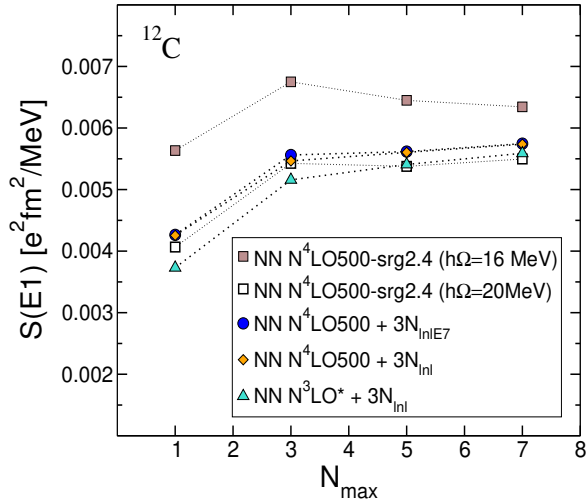}
 		\caption{Calculated sum of $E1$ matrix elements ($S(E1)$ in Eq.~\ref{eq:SE1}) using the {\sc NCSM} and different $\chi$EFT interactions
 		over all intermediate states $\lvert J_n\rangle = \lvert 1^-_{_{GDR}}\rangle$ connecting both $\lvert 0^+_{_1}\rangle$ and $\lvert 2^+_{_1}\rangle$ states.
 		The frequencies used for the N$^4$LO interactions were $\hbar\Omega=$18 MeV while for N$^3$LO $\hbar\Omega=$20 MeV unless otherwise specified in the legend.}
 		\label{fig:SE1calc}
 	\end{center}
\end{figure}

In the present  work $\kappa(2_{_1}^+)$ has been calculated using the {\it ab~initio} {\sc NCSM} method with the Similarity Renormalization Group ({\sc SRG}) evolved~\cite{bogner2007similarity} non-local $\chi$EFT NN interaction, NN N$^4$LO500--srg2.4~\cite{entem2015peripheral,entem2017high}, employed in our previous study~\cite{raju2018reorientation}, as well as with three $\chi$EFT interactions, also {\sc SRG} evolved, that include 3N forces; namely, {\sc NN} N$^3$LO$^*$ + {\sc 3N}$_{lnl}$~\cite{entem2003accurate,soma2020novel}, {\sc NN} N$^4$LO + {\sc 3N}$_{lnl}$~\cite{entem2017high,gysbers2019discrepancy}, and the recently developed {\sc NN} N$^4$LO + {\sc 3N}$^*_{lnl}$ interaction~\cite{entem2017high,kravvaris2023ab,girlanda2011subleading}, which contains a sub leading {\sc 3N} contact term ({\sc E7}) that enhances the strength of the spin-orbit interaction~\cite{girlanda2011subleading}. The latter is the interaction that better predicts the excitation energy, $E(2^+_{_1})=4.1$ MeV (4.44 MeV experimentally).
The {\sc NCSM} calculations have been performed in basis spaces ranging from $N_{_{max}} = 0/1$ to $N_{_{max}} = 6/7$, where the odd $N_{_{max}}$ spaces were used for
the intermediate 1$^-$ states and even $N_{_{max}}$ for the 0$^+_1$ and 2$^+_1$ states, with the odd $N_{_{max}}$ calculation matched with the even $N_{_{max}}-1$ calculation.
We implemented the Lanczos continued fraction algorithm ({\sc LA}) method~\cite{Marchhisio2003} to perform exactly the summation over the intermediate 1$^-_{_{GDR}}$ states in Eq.~\ref{eq:SE1}.
The {\sc NCSM-LA} method allows to account for the gamma strength well above 30 MeV that was the limit we reached in the previous {\sc NCSM} calculations with the intermediate states explicitly computed, e.g., 28 states in $N_{_{max}} = 6/7$~\cite{raju2018reorientation}.  Only in the $N_{_{max}} = 0/1$ space, we can calculate all the 1$^-$ states (248) and perform the sum~\ref{eq:SE1} explicitly.
The result then serves as a benchmark for the {\sc LA} method. We note that no effective charges are used in the {\sc NCSM} calculations.

In Fig.~\ref{fig:SE1calc}, we present the dependence of  $S(E1)$ (\ref{eq:SE1}) on the basis size $N_{_{max}}$ for all the used chiral interactions.
We find a good convergence that allows a robust inference of $S(E1)=0.0060(5)$ e$^2$ fm$^2$/MeV. This result is slightly higher but still consistent with the calculation reported in Ref.~\cite{barke1989investigation}. We note that a good convergence of $S(E1)$ is consistent with other {\sc NCSM-LA} calculations of various observables that involve a Green’s function, including anapole and electric dipole moments~\cite{hao2020nuclear,froese2021ab,gennari2024textit}. This is in contrast to a slow convergence of observables sensitive to the wave function tail such as the quadrupole moments and radii. Consequently, in order to obtain a robust prediction for $\kappa(2^+)$, we use the experimental value of the matrix element $\langle 0^+_{_1}\mid\mid \hat{E2} \mid\mid 2^+_{_1}\rangle=6.247(63)$ efm$^2$
(or $B(E2; 0^+_{_1} \rightarrow 2^+_{_1}) = 39.0(8)$ e$^2$ fm$^4$)  in Eq.~\ref{eq:extd_kappa},  yielding $\kappa(2_{_1}^+) =  1.2(1)$. We also computed $\kappa(0_{_1}^+)= 1.8(1)$ within the {\sc NCSM-LA} approach, which is consistent with the experimental value of $\kappa(0^+_{_1})=1.6(2)$ arising from Fuller’s corrections~\cite{fuller1985photonuclear} to the overestimated photo-absorption cross-section data by Ahrens and co-workers~\cite{ahrens1975total}.

\begin{figure*}[!ht]
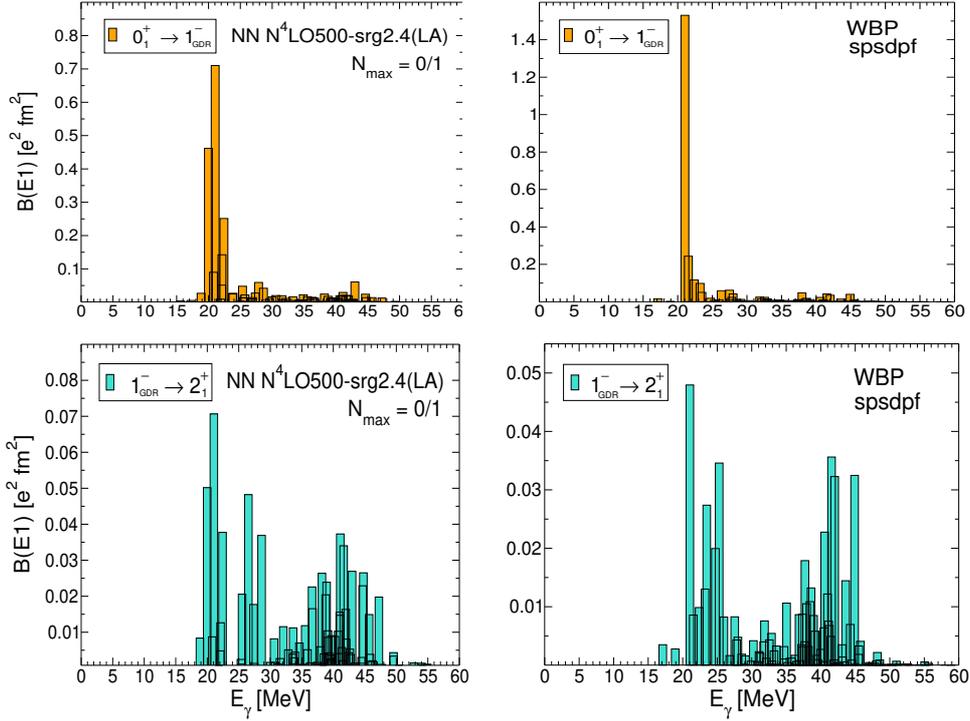

\begin{center}
\includegraphics[width=6.1cm,height=4.2cm,angle=-0]{BE1_C12_n4lo500-srg2.4_Nmax0_Nmax1_LA.eps} \hspace{0.3cm}
\includegraphics[width=6.cm,height=4.2cm,angle=-0]{BE1_C12_WBP2.eps}\\
\vspace{0.3cm}
\includegraphics[width=6.1cm,height=5.cm,angle=-0]{BE1_C12_n4lo500-srg2.4_Nmax0_Nmax1_LA2.eps} \hspace{0.3cm}
\includegraphics[width=6.1cm,height=5.cm,angle=-0]{BE1_C12_WBP.eps}
\caption{The $B(E1)$ strengths calculated using the NN N$^4$LO500-srg2.4  with $N_{_{max}}= 0/1$ (left panels) and WBP (right panels) interactions for the transitions $0_{_1}^+ \rightarrow 1_{n}^-$ (top) and $1_{n}^-\rightarrow 2_{_1}^+$ (bottom).}
\label{fig:BE1_plots}
\end{center}
\end{figure*}

The present $\kappa(2_{_1}^+)$ value is lower than that obtained in Ref.~\cite{raju2018reorientation} for two reasons: (i) All 1$^-_{_{GDR}}$ states have been included to compute $S(E1)$ in the present study by the
{\sc LA} method contrary to just 28 states computed explicitly in ~\cite{raju2018reorientation}. It turns out that the higher-lying states above the main {\sc GDR} peak contribute destructively to $S(E1)$. This is consistent with calculations reported in Ref.~\cite{barke1989investigation}.
We have verified this explicitly in the $N_{_{max}} = 0/1$ calculations shown in the left panels of Fig.~\ref{fig:BE1_plots} (note that only the $B(E1)$
values connecting the 0$^+_{_1}$ and 2$^+_{_1}$ states to the 1$^-_{_{GDR}}$ states are shown, not the $S(E1)$ contributing terms).
(ii) We use the experimental value of $\langle 0^+_{_1}\mid\mid \hat{E2} \mid\mid 2^+_{_1}\rangle$, which is larger than the {\sc NCSM} ones in the $N_{_{max}}$ spaces employed here
and in Ref.~\cite{raju2018reorientation}.  The obtained $\kappa(2_{_1}^+)$ value is in line with what is found in other self-conjugate nuclei in the $psd$ shells~\cite{Orce2024,orce2024global}.
The left panels of Fig.~\ref{fig:BE1_plots}
also reveal the weak coupling  between the 2$^+_{_1}$ and 1$^-_{_{GDR}}$  states.

%

\begin{figure*}[!ht]
\begin{center}
\includegraphics[width=6.5cm,height=5.5cm,angle=0]{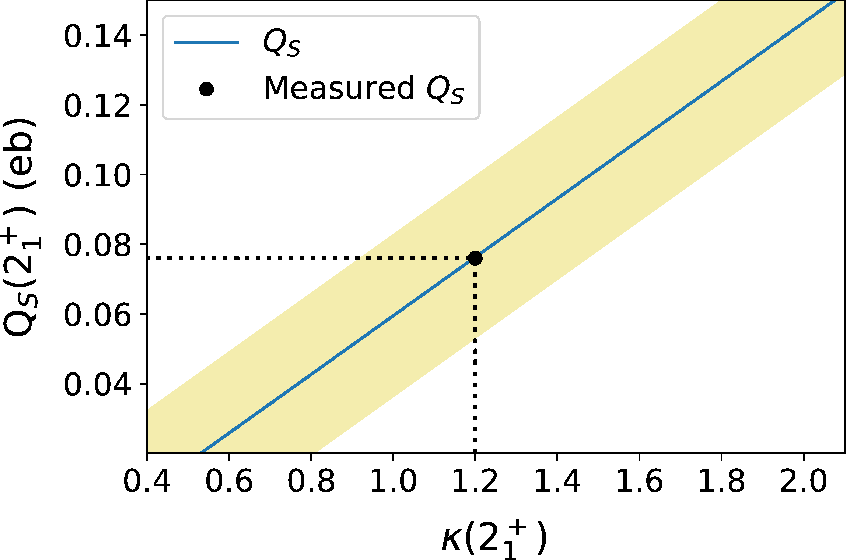}
\hspace{0.2cm}
\includegraphics[width=6.2cm,height=5.4cm,angle=-0]{Xicurve2.eps}
\caption{(Left panel) General trend of the extracted Q$_{_S}(2_{_1}^+)$ by fitting population probabilities from {\sc GOSIA} as a function of $\kappa$. The shaded band represent the uncertainty region. (Right panel) The $X_i$ curve obtained from the least squares search and the red data point with an error bar corresponds to the measured $Q_{_S}(2_{_1}^+)$.}
\label{fig:chi_curveq}
\end{center}
\end{figure*}

Further {\sc SM} calculations using the phenomenological {\sc WBP} interaction~\cite{warburton1992effective} in the $spsdpf$ model space
with effective charges $e_p=1.275$ and $e_n=0.275$~\cite{bassel19820+,yuan2012}
yields $B(E2; 0^+_{_1}\rightarrow 2^+_{_1}) = 0.004254$~e$^2$b$^2$,
in agreement with experiment~\cite{raman2001transition,pritychenko2016tables}, and $Q_{_S}(2_{_1}^+)$ = $0.0623$~eb.
Additionally, the {\sc WBP} interaction predicts a slightly smaller  $\kappa(2_{_1}^+) = 0.9$ and larger $\kappa(0^+_{_1}) = 3.2$ values.
As shown in the right panels of Fig.~\ref{fig:BE1_plots}, the distribution of $E1$ strength up is less fragmented than the {\sc NCSM}
calculation and concentrated at 21 MeV for the $0^+_{_1}\rightarrow 1_{_{GDR}}^-$ calculation, which is lower than the experimental $E_{_{GDR}}\approx24$~MeV~\cite{fuller1985photonuclear}. We note that the
{\sc WBP} interaction estimates a  binding energy of 102.36 MeV for the ground states, that is $\approx 10$ MeV larger than the measured 92.16 MeV~\cite{wang2021ame}.

\section{Results and Discussion}

Empirically, the $Q_{_S}(2_{_1}^+)$ value is determined utilizing population probabilities. In a combined analysis using the sum of the populations of all angles,
we performed a least squares search by varying $\langle 2^+_{_1} \mid \mid \hat{E2} \mid \mid 2^+_{_1} \rangle$ until
the population probabilities calculated by {\sc GOSIA} converge with the experimental value determined by Eq.~\ref{eq:P_exp}.
The least squares is described by 
\begin{equation}
  X_i = \frac{\big[P(^{12}\mbox{C};2_{_1}^+)-P_{i}(\langle 2^+_{_1} \mid \mid \hat{E2} \mid \mid 2^+_{_1} \rangle)\big]^2}{\sigma_{_P}^2},  
  \label{eq:chi}
\end{equation}
where $\sigma_{_P}$ is the uncertainty of P($^{12}$C;2$_{1}^+$) and P$_i$ is the population probability from {\sc GOSIA}, whose magnitude varies with  $\langle 2^+_{_1} \mid \mid \hat{E2} \mid \mid 2^+_{_1} \rangle$.
Convergence of the probabilities is reached at a global minimum $X_{min} = 0$, where P($^{12}$C;2$_{1}^+$) $=$ P$_i$. The right panel of Fig.~\ref{fig:chi_curveq} presents the $X_i$ curve of our analysis as a function of the assumed Q$_{_S}(2_{_1}^+)$ value. The $X_{min}$ corresponds to Q$_{_S}(2_{_1}^+) = $~0.076(30) eb, which is the quoted measurement in this work. Uncertainty of the latter stems from the statistical uncertainty determined by finding the value of Q$_{_S}(2_{_1}^+)$ corresponding to P($^{12}$C;2$_{1}^+$) $\pm$ $\sigma_{_P}$ in the GOSIA analysis, and phenomenological effects, i.e.,
high-lying $2_{_2}^+$ and $0_{_2}^+$ states in $^{12}$C, quantal effects and the $\kappa(2_{_1}^+)$ value. The $2_{_2}^+$ and $0_{_2}^+$ states have negligible effect on our measurement $\sim 0.4\times10^{-3}$ eb. In order to ascertain the confidence levels of the {\sc SCA} in support of our argument regarding the ``safety'' of the experiment, quantum effects whose magnitude is proportional to $1/\eta = 1/38$ ($\sim 2.6\%$), must be added to the uncertainty. Another vital parameter is the fixed transitional matrix element in GOSIA;
$\langle 0^+_{_1}\mid\mid \hat{E2} \mid\mid 2^+_{_1}\rangle=0.06247(63)$ eb. Lastly, the $\kappa(2_{_1}^+) = 1.2(1)$ value taken as input in {\sc GOSIA}, is the only model dependent parameter.
Therefore, we have investigated the correlation of $\kappa(2^+_{_1})$ and Q$_{_S}(2_{_1}^+)$, as shown in the left panel of Fig.~\ref{fig:chi_curveq}.
It is clear that increasing $\kappa(2_{_1}^+)$ induces a larger  oblate deformation.
From the observed correlation, the uncertainty on $\kappa(2_{_1}^+)$ affects Q$_{_S}(2_{_1}^+)$ by 0.003 eb. A summary of our results is presented in Table~\ref{Tbl:results}.


\begin{table}[!ht]
 \centering
 \vspace{0cm}
 \begin{tabular}{c c c c c }
     \hline \hline
     Model              &  B(E2;$0^+_{_1} \rightarrow 2^+_{_1}$)    & $Q_{_S}(2_{_1}^+)$   & $\kappa(0_{_1}^+)$  & $\kappa(2_{_1}^+)$   \\
                        &           e$^2$b$^2$                      &     eb               &                     &               \\
     \hline
 Shell Model            &           0.00425                         &   0.0623             &  3.23               &   0.9     \\

 {\sc PGCM}             &           0.00406                         &   0.057              &  --                 &   --   \\

 {\sc NCSM-LA}          &  --                                       &  --                  &  1.8(1)             &   1.2(1)     \\

 Rigid rotor            &  --                                       & 0.0566(6)            &   --                &   -- \\

 \hline
 Experiment             &           0.00390(8)~\cite{pritychenko2016tables,d2020precision}                      & 0.076(30)$^*$        & 1.6(2)~\cite{fuller1985photonuclear}   &  --  \\
 \hline \hline

 \end{tabular}
 \caption{Theoretical model predictions of dipole and quadrupole parameters compared with experiment data.
 Calculations were performed using the {\sc WBP} interaction~\cite{warburton1992effective} with the shell model,
 the N$^2$LO EM1.8/2.0 interaction~\cite{hebler} with the {\sc PGCM}, and four chiral interactions (NN N$^4$LO500--srg2.4~\cite{entem2015peripheral,entem2017high},
 {\sc NN} N$^3$LO$^*$ + {\sc 3N}$_{lnl}$~\cite{entem2003accurate,soma2020novel}, {\sc NN} N$^4$LO + {\sc 3N}$_{lnl}$~\cite{entem2017high,gysbers2019discrepancy} and the recently developed {\sc NN} N$^4$LO + {\sc 3N}$^*_{lnl}$ interaction~\cite{entem2017high,kravvaris2023ab,girlanda2011subleading} with the {\sc NCSM-LA}.
 The {\sc NCSM} values for B(E2;$0^+_{_1} \rightarrow 2^+_{_1}$)  and  $Q_{_S}(2_{_1}^+)$ are not shown as they are far from convergence in the accessible $N_{_{max}}$ spaces, see Fig.~\ref{QsvsNmax}.
The rigid rotor value is given by $Q_{_S}(2^+_{_1})=-\frac{2}{7}\left(\frac{16\pi}{5}~B(E2; 0^+_1\rightarrow 2^+_1\right)^{1/2}=0.9059 ~B(E2; 0^+_1\rightarrow 2^+_1)^{1/2}$~\cite{BohrMottelson,deshalit1990theoretical}.  An asterisk indicates the $Q_{_S}(2_{_1}^+)$ value arising from this work.}
 \label{Tbl:results}
\end{table}

As shown in the right panel of Fig.~\ref{fig:chi_curveq}, our result is within errors in concurrence with the large oblate deformation determined  in previous {\sc RE}  measurements~\cite{vermeer1983electric,raju2018reorientation,saiz2021coulomb}.
It is worth noting the analogous experiment by Vermeer and co-workers~\cite{vermeer1983electric}, where a reported value of $Q_{_S}(2_{_1}^+) = 0.06(3)~eb$ was determined for $\kappa(2_{_1}^+)=0.68$
(i.e., $\kappa(2_{_1}^+)=1$  using Levinger's $\sigma_{_{-2}}=3.5 \kappa~ A^{5/3}$ $\upmu$b/MeV superseded formula~\cite{levinger1957migdal}, instead of Eq.~\ref{eq:238}). 
A re-analysis of this prior work using the currently more accurate weighted average, $B(E2; 0^+_{_1} \rightarrow 2_{_1}^+) = 0.00390(8)$ e$^2$b$^2$ vs $B(E2; 0^+_{_1} \rightarrow 2_{_1}^+) = 0.00388(22)$ e$^2$b$^2$~\cite{vermeer1983electric} --- which includes all previous measurements~\cite{pritychenko2016tables,d2020precision} --- and the newly calculated $\kappa(2_{_1}^+) =1.2(1)$ yields  a larger and more precise value of $Q_{_S}(2_{_1}^+)=0.103(20)~eb$. Additionally, a recent result by Saiz-Lomas and co-workers~\cite{saiz2021coulomb} yields a larger $Q_{_S}(2_{_1}^+) = 0.093^{+3.5}_{-3.8}$~eb using $\kappa(2_{_1}^+) = 1.0(2)$ and the high-precision lifetime measurement~\cite{d2020precision}, in agreement with the present work. Combining this and previous work, a weighted average of $Q_{_S}(2_{_1}^+) = 0.090(14)$~eb is determined, which clearly challenges modern nuclear theory.



\begin{figure}[!ht]
\begin{center}
\includegraphics[width=7.6cm,height=6.5cm,angle=-0]{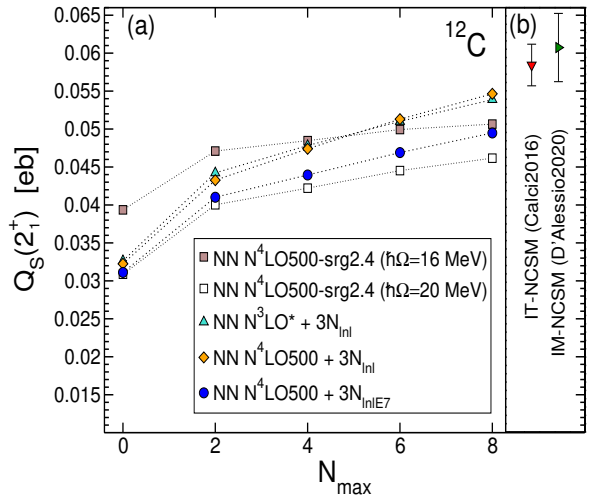}
\caption{Panel (a) shows $Q_{_S}(2_{_1}^+)$ values calculated with the {\sc NCSM} as a function of $N_{_{max}}$ using four different $\chi$EFT  interactions.
Panel (b) presents extrapolated $Q_{_S}(2_{_1}^+)$ values obtained by Calci {\it et al}.~\cite{calci2016sensitivities} and D'Alessio {\it et al}.~\cite{d2020precision} from a B(E2;$2^+_{_1} \rightarrow 0^+_{_1}$) -- $Q_{_S}(2_{_1}^+)$ correlation.}
\label{QsvsNmax}
\end{center}
\end{figure}

In panel (a) of Fig.~\ref{QsvsNmax}, we present {\sc NCSM} calculations of $Q_{_S}(2_{_1}^+)$ obtained with the four chiral interactions used to compute $S(E1)$ (shown in Fig.~\ref{fig:SE1calc}).
The calculations have been performed in models spaces up to $N_{_{max}}=8$ without any further truncation. For the NN N4LO500–srg2.4, results are shown for two frequencies, $\hbar\Omega =$ 16 MeV
(upper line) and $\hbar\Omega=$ 20 MeV (lower line) yielding $Q_{_S}(2_{_1}^+) \approx 0.05$ eb. However, the $Q_{_S}(2_{_1}^+)$ values obtained with the more realistic chiral NN+3N interactions are clearly far from convergence. In panel (b), we show $Q_{_S}(2_{_1}^+)$ values calculated from a correlation
of $Q_{_S}(2_{_1}^+)$ and B(E2;$2^+_{_1} \rightarrow 0^+_{_1}$) values by Calci {\it et al.}~\cite{calci2016sensitivities} and D'Alessio {\it et al.}~\cite{d2020precision} using different chiral interactions and  momentum cut-off parameter, $\varLambda$, yielding results of $Q_{_S}(2_{_1}^+) \approx 0.06$ eb.
These  studies~\cite{d2020precision,calci2016sensitivities} also reveal that $Q_{_S}(2_{_1}^+)$ is insensitive to $\varLambda$.

The most likely explanation for the larger oblate deformation found in this work concerns $\alpha$-cluster effects that are not properly included in the {\sc NCSM}.
In fact, the degree of  $\alpha$ clustering in $^{12}$C has been investigated within the Fermionic Molecular Dynamics (FMD) and  microscopic
$\alpha$-cluster models~\cite{chernykh2007structure},  showing that the ground and 2$^+_{_1}$ states present considerable $\alpha$-cluster triangle admixtures
of 52\% and 67\%, respectively.
Further, {\sc GDR} states mainly arise from coherent 1-particle-1-hole (1p-1h) excitations, and  $\alpha$ clustering may not affect the {\sc NCSM} calculations of $E1$ matrix elements
in order to extract the dipole polarizability.
However, an interesting work by He and collaborators shows that $\alpha$ configurations may affect nuclear collective motion, specifically the
{\sc GDR} excitation, giving rise to a highly fragmented {\sc GDR} spectrum~\cite{he2014giant}.
The explicit account of deformation properties with chiral interaction can be efficiently probed via the projected generator coordinate method (PGCM)~\cite{frosini2022multi,marevic2019cluster}.
This method mixing mean-field solutions with different deformations and orientations has been shown to incorporate efficiently long-range correlations.
A {\sc PGCM} calculation restricted to axially symmetric shapes and performed at $N_{{max}}=11$ with EM1.8/2.0 Chiral interaction~\cite{hebler} yields E($2_{_1}^+$) = 4.2 MeV,
$B(E2; 0^+_{_1} \rightarrow 2_{_1}^+) = 0.00406$ e$^2$b$^2$ and $Q_{_S}(2_{_1}^+) = 0.057$ eb.
While the first two observables are well reproduced, the too low spectroscopic quadrupole moment calls for triaxial configurations to account
for triangle $\alpha$ structures in $^{12}$C ground-state band.

\section{Conclusions}

In summary, we present a measurement of $Q_{_S}(2_{_1}^+)=+0.076(30)$ eb in $^{12}$C by Coulomb excitation using the $^{208}$Pb($^{12}$C,$^{12}$C$^*$)$^{208}$Pb$^*$ reaction at a safe energy of 56 MeV and the {\sc Q3D} magnetic spectrograph positioned at angles $80^\circ$, $85^\circ$ and $90^\circ$. The analysis was carried out using population probabilities from {\sc GOSIA}, assuming $\kappa(2_{_1}^+) = 1.2(1)$ determined from {\sc NCSM} (using four chiral interactions) and the $B(E2; 0^+_{_1} \rightarrow 2_{_1}^+)$ = 0.00390(8) e$^2$b$^2$ weighted value.
Combined with previous work a large oblate deformation of $Q_{_S}(2_{_1}^+)=+0.090(14)$ eb is determined, which challenges the rigid-rotor model at the  1.3$\sigma$ level
as well as modern state-of-the-art {\it ab initio} calculations.
Larger and untruncated  harmonic oscillator basis sizes ($N_{_{max}}>10$) (or explicit introduction of
$\alpha-$clusters~\cite{Navratil2016,kravvaris2024ab}) and possibly higher leading-order $\chi$EFT interactions~\cite{epelbaum2012structure,stroberg2022systematics,henderson2022coulomb} are probably required to include $\alpha$-clustering effects,
and, particularly for $^{12}$C, the explicit inclusion of triaxiality, for convergence of $E2$ properties.
The later is a problem faced by the $\chi$EFT community with a particular challenge of determination of low-energy constants for the $3N$ force at 5$^{th}$  order~\cite{girlanda2011subleading}.
Moreover, this work sheds light onto the impact of large $\kappa$ values on $E2$ properties, particularly in low mass nuclei, and further calls for dedicated Coulomb-excitation measurements of the nuclear dipole polarizability in order to disentangle the {\sc RE} and $E1$-polarizability effects in even-even nuclei, a research ground that remains practically untouched.


\section*{Declaration of competing interest}
There are no known competing financial interests or personal relationships that could have appeared to influence the work reported in this paper.

\section*{Acknowledgements}
This work was supported by the National Research Foundation (NRF) of South Africa under Grant 93500 and by the NSERC Grant No. SAPIN-2022-00019.
TRIUMF receives federal funding via a contribution agreement with the National Research Council of Canada. Computing support came from an INCITE Award on the Summit and Frontier supercomputers of the Oak Ridge Leadership Computing Facility (OLCF) at ORNL, LLNL LC, and from the Digital Research Alliance of Canada.
We also acknowledge  the CEA-SINET project and the CCRT HPC resource (TOPAZE supercomputer) for PGCM calculations.
We finally thank
the late B. Singh for providing relevant information and  M. Zieli\'nska
for contributions during and after the experiment.

\bibliographystyle{elsarticle-num} 
\bibliography{ref}






\end{document}